\documentclass[aps,showpacs,prl,twocolumn]{revtex4}

\begin{document}


\title {Large $N_c$ Continuum Reduction and the Thermodynamics of QCD}

\author{Thomas D. Cohen}
\email{cohen@physics.umd.edu}

\affiliation{Department of Physics, University of Maryland,
College Park, MD 20742-4111}

\begin{abstract}
It is noted that if large $N_c$ continuum reduction applies to an
observable, then that observable is independent of temperature
for all temperatures below some critical value.  This fact, plus
the fact that mesons and glueballs are weakly interacting at
large $N_c$ is used as the basis for a derivation of large $N_c$
continuum reduction for the chiral condensate. The structure of
this derivation is quite general and can be extended to a wide
class of observables.
\end{abstract}


\maketitle

One of the most intriguing  recent developments in QCD is the
prospect of large $N_c$ continuum
reduction\cite{lncr1,lncr2,lncr3,lncr4}. Underlying this is the
two decades old idea of Eguchi and Kawai (EK) that as $N_c$, the
number of degrees of freedom in QCD, goes to infinity, the
relevant spatial size in the dynamics goes to zero\cite{EK}. This
was originally formulated on the lattice and it was shown that the
Wilson loop was computable on a single plaquette, from which it
is deduced that the free energy density is computable on a single
plaquette. Unfortunately, although the EK analysis is formally
correct, it has been of limited practical significance there is a
phase transition separating the EK reduced theory from the
physical point as one approaches the continuum limit\cite{BHN}.
However, recently it was suggested that while the continuum
theory  cannot be reduced to a point---it can be reduced to a
Euclidean box of finite size ({\it in terms of physical units})
while keeping the exact large $N_c$ infinite volume
result\cite{lncr1,lncr2,lncr3,lncr4}. The guts of this argument
is that provided the box size is larger than the critical size
for which the phase transition occurs, the EK reduction works and
one can simultaneously avoid the phase transition and thus
connect with the continuum limit. Numerical simulations in three
and four dimensions\cite{lncr1,lncr2,lncr3,lncr4} provide strong,
but not compelling, evidence that this scenario is correct.  The
phenomena is referred to as large $N_c$ continuum reduction.

It has also been conjectured that the same phenomena of large
$N_c$ continuum reduction occurs for the chiral condensate:
namely, the chiral condensate as computed in a finite box at
large $N_c$ becomes independent of box size for boxes larger than
some critical value\cite{cc}.  The conjecture is motivated by
consideration of random matrix theory (RMT) which is believed to
become universal for the lowest modes of the Dirac operator; this
presumed universal behavior is  in terms of the natural scaled
variable in RMT, $z_k=\lambda_k N_c L^4 \Sigma$ where $\sigma$ is
the chiral condensate and $\lambda_k$ is the $k$th eigenvalue of
the Dirac operator.  One expects the same universal behavior by
taking $L \rightarrow \infty$ or $N_c \rightarrow \infty$.

This conjecture may seem paradoxical: it is well known that
spontaneous symmetry breaking cannot occur for systems of spatial
extent. The resolution to this paradox is simple: a necessary
(though not sufficient) condition for spontaneous symmetry
breaking is not infinite spatial size but rather an infinite
number degrees of freedom (which, of course, occurs for a system
of infinite spatial extent).  There can be an infinite number of
degrees of freedom in a system of finite spatial extent if one
has an infinite number of internal degrees of freedom; this is
precisely what happens for QCD as $N_c \rightarrow \infty$. There
is numerical evidence that the chiral condensate can be computed
on a finite box by an extrapolation to infinite $N_c$; moreover,
the evidence is consistent with the possibility that the value of
the condensate is independent of the box size beyond some
critical size \cite{cc}. If the continuum reduction holds, then
one can do reliable large $N_c$ simulations on relatively size
boxes leading to relatively inexpensive computations.  This
prospect is exciting;  reliable and converged lattice QCD results
may be available in the near term---albeit for a colorful world.

The present letter has two main purposes: to point out that the
continuum reduction has very strong implications for the
phenomenology of large $N_c$ QCD at nonzero temperature and to
provide a basis in well-known large $N_c$ phenomenology for
understanding why continuum reduction must occur for the chiral
condensate (and for the expectation values of a wide class of
other operators). These two issues are intimately related.

First consider the phenomenological implications for large $N_c$
QCD at $T \ne 0$.  The central result is that any quantity which
is subject to the continuum reduction at large $N_c$ has the
property that for any temperature less than some critical
temperature $T_c$, the quantity is unchanged from its $T=0$
value. This result is not exact at finite $N_c$ but any
deviations from the $T=0$ are subleading:
\begin{equation}
\frac{Q(T) -Q(0)}{Q(0)} \sim 1/N_c \; \; {\rm for} \; \; T<T_c
\label{phen}
\end{equation}
where $Q(T)$ is some quantity subject to large $N_c$ continuum
reduction. Thus, for example, if the chiral condensate is subject
to the continuum reduction (as conjectured in ref.~\cite{cc}) then
in a large $N_c$ world as the temperature of the system as
increased from zero, the chiral condensate will remain unchanged
from its $T=0$ value until the temperature reaches a critical
point.

The physical basis is quite simple. The logic of large $N_c$
continuum reduction implies that when the Euclidean space box size
is larger than the critical size, the results at large $N_c$ do
not depend on the size {\it or shape} of the box: if all of the
sides of the box are larger than the critical length ($L_c$),
then the quantity  does not depend further on the lengths of the
sides.  Thus for a box with three of the sides of infinite extent
and the fourth of finite size (greater then $L_c$), then a
quantity subject to continuum reduction does not depend on the
length of this fourth side. However, Euclidean field theory with
one side of finite length and standard boundary conditions
(periodic for bosons, anti-periodic for fermions) {\it is} just
finite temperature field theory  with $T=1/L$ (where $L$ is the
length of this side).  Thus, any quantity for which large $N_c$
continuum reduction holds will be independent of temperature for
$T<T_c$ where $T_c=1/L_c$ up to $1/N_c$ corrections.

This behavior is consistent with what is known about the
temperature dependence of the chiral condensate. Consider the
chiral condensate at fixed low temperature in the combined large
$N_c$ and chiral limits.  For simplicity, consider the case where
the chiral limit is taken prior to the large $N_c$ limit; this
avoids ambiguity about defining the chiral condensate by limiting
the sensitivity to the ultraviolet.  Moreover, with this ordering
chiral perturbation theory at finite temperature can be used to
compute an analytic expression for the variation of the chiral
condensate with $T$ (at zero quark mass) \cite{chpt}. Up to order
$T^6$ the result is:
\begin{equation}
\frac{\langle \overline{q} q \rangle_{T}}{\; \; \, \,\langle
\overline{q} q \rangle_{T=0}} = 1 - \frac{T^2}{8 f_\pi^2} -
\frac{T^4}{384 f_\pi^4} + \frac{T^6}{288 f_\pi^6} \log \left(
\frac{T}{\Lambda_q} \right ) + {\cal O}(T^8)  \label{xpt1}
\end{equation}
where $\langle \overline{q} q \rangle_{T}$ is the chiral
condensate at temperature $T$, $f_\pi$ is the pion decay constant
(evaluated in the chiral limit) and $\Lambda_q$ is a parameter
which depends on higher-order counter terms in the chiral
lagrangian.  Recall the standard $N_c$ scaling of $f_\pi$:$ f_\pi
\sim N_c^{1/2} $.  Using this scaling and taking temperatures of
order $N_c^0$, and combining it with eq.~(\ref{xpt1}) yields the
result that
 \begin{equation}
\lim_{N_c \rightarrow \infty} \frac{ \langle \overline{q} q
\rangle_{T}}{\; \; \, \, \langle \overline{q} q \rangle_{T=0}}=1+
{\cal O}(T^8) \; . \label{chp}
\end{equation}
There is no variation of the chiral condensate with temperature in
a large $N_c$ world up to the order calculated, as is required by
large $N_c$ continuum reduction.

Equation (\ref{chp}) is highly suggestive and its important to
understand whether it holds generally. Fortunately, it is possible
to show on very general physical grounds that the chiral
condensate is independent of temperature at sufficiently low
temperatures in a large $N_c$ world. Obviously, this is of
phenomenological interest (at least in a large $N_c$ world). In
addition, this observation can be used as the physical  basis of a
demonstration that the phenomena of continuum large $N_c$
reduction applies to the chiral condensate. Moreover, the logic
underlying the derivation can be generalized to the expectation
values of a wide class Lorentz scalar local operator and not just
to the this chiral condensate.

The key to this general derivation is the fact that standard `t
Hooft counting\cite{'tH} implies that at large $N_c$, mesons and
glueballs have masses of order $N_c^0$ and are weakly interacting:
$V_{\rm meson-meson} \sim N_c^{-1}$, $V_{\rm  meson-glueball} \sim
N_c^{-2}$ and $V_{\rm glueball-glueball}  \sim  N_c^{-2}$ where
$V$ indicates the strength of the two-body interaction
(four-point function). Now consider an extended system in which
the density of mesons and glueballs is of order $N_c^0$.  In such
a system the thermodynamics will be simply that of a gas
non-interacting mesons and glueballs plus corrections which
vanish at large $N_c$.  Thus, at temperature of order $N_c^0$,
the density of a given species, $s$, of a meson or glueball is
given by the standard expression for noninteracting bosons:
\begin{widetext}
\begin{equation}
\rho_s(T)  =  d_s \int \frac{{\rm d}^3 k }{(2 \pi)^3} \,
\rho(T,m_s,k) \, \left ( 1 + {\cal O}(N_c^{-1}) \right ) \; \;
\\ {\rm with} \; \; \rho(T,m_s,k)\equiv  \frac{1}{ e^{ \sqrt{k^2 + m_s^2} / T} -1}\label{den}
\end{equation}
 where $d_s \equiv (2 J_s+1)(2 I_s +1)$ is the
degeneracy  factor for a species with angular momentum $J_s$ and
isospin $I_s$. Note in terms of $N_c$ counting $\rho_s(t) \sim
N_c^0$. The free energy density as measured relative to its zero
temperature value is computed in a similar manner and is given by
\begin{equation}
{\cal G}_r = - T \sum_s d_s \int \frac{{\rm d}^3 k }{(2 \pi)^3}
\log \left (1- e^{- \sqrt{k^2 + m_s^2} / T} \right )\left ( 1 +
{\cal O}(N_c^{-1} ) \right ) \label{G}
\end{equation}
\end{widetext} where the summation is over species of boson.  The
subscript $r$ indicates that the free energy is measured relative
to its $T=0$ value: ${\cal G}_r(T)={\cal G}(T)-{\cal G}(0)$.
Normally, by convention one sets ${\cal G}(0)$ to zero and this
distinction is unimportant.  However, in the present context we
will differentiate with respect to an external source, such as the
quark mass, and ${\cal G}(0)$ has nonvanishing dependence on these
sources.

Before proceeding it is useful to recall that strength of the
interaction between a meson or a glueball and a baryon is of
order unity in $N_c$ counting and thus is not weakly
interacting.  So, if the system has a baryon density of order
unity, eqs.~(\ref{den}) and (\ref{G}) will be spoiled. However,
the baryon mass, unlike the meson or glueball masses, is order
$N_c$ \cite{Wit} and hence, for temperature of order $N_c^0$, the
density of baryons will be of order $e^{- N_c}$ and is negligible
in the large $N_c$ limit.

Consider the chiral condensate, {\it i.e.},  the expectation value
of $\overline{q}q$ where both quarks are understood to be at the
same space-time position. Of course, this operator is not well
defined without specifying some prescription for how to regulate
it at large momentum. But, as we will see for the present purpose,
the details of this prescription turns out to be largely
irrelevant. For the moment let us neglect this issue and proceed
formally.  To simplify issues associated with renormalization it
is useful to discuss the RG invariant operator $m \overline{q}q$
(where $m$ is the quark mass) rather than $\overline{q}q$ itself.
From the standard form of the grand partition function as a
Euclidean space functional integral it is immediately apparent
that
\begin{equation}
m \langle \overline{q}q \rangle = m \frac{T}{V} \frac{d \log
(Z)}{d m} \label{ccz}
\end{equation}
where the system is confined to a three-space volume $V$ (taken
to infinity in the thermodynamic limit) and is of length
$\beta=1/T$ in the time direction; $m$ is the quark mass.
Defining the free energy density in the usual way as ${\cal G}=
-\frac{T}{V}\log (Z)$ it is apparent that $\langle \overline{q}q
\rangle= \frac{d {\cal G} }{d m}$. Combining this with eqs.
(\ref{den}) and(\ref{G}) along with the definition of ${\cal G}_r$
yields:\begin{widetext}
\begin{equation}
m \langle \overline{q}q\rangle_T - m \langle \overline{q}q
\rangle_{T=0} = \sum_s d_s \sigma_s  \int \frac{{\rm d}^3 k }{(2
\pi)^3} \, \rho(T,m_s,k) \, \frac{ \sigma_s m_s}{\sqrt{k^2 +
m_s^2}} \left ( 1 + {\cal O}(N_c^{-1})\right ) \label{simple}
\end{equation}
\end{widetext}
 where $\sigma_s$ the sigma commutator for the
species $s$ and is defined by $\sigma_s \equiv m \frac{d m_s}{d
m}$.

The key point for the purposes here is the $N_c$ scaling.  By
conventional large $N_c$ scaling arguments\cite{'tH,Wit} it is
straightforward to see that
\begin{eqnarray} \sigma_s & \sim &
N_c^0 \; \;{\rm for} \; s \in \{{\rm mesons}\}
\nonumber \\
\sigma_s & \sim & N_c^{-1} \; \;{\rm for} \; \; s \in \{{\rm
glueballs}\} \nonumber \\
 \rho(T,m_s,k) & \sim &
N_c^0\; \;{\rm for} \; \; T \sim N_c^0 \nonumber \\
m \langle \overline{q}q \rangle_{T=0} &\sim & N_c^1 \,
.\label{sigscale}
\end{eqnarray}
Combining these scaling rules with eq.~(\ref{simple}) yields,
\begin{equation}
\frac{m \langle \overline{q}q\rangle_T - m \langle \overline{q}q
\rangle_{T=0}}{ m \langle \overline{q}q \rangle_{T=0}} \sim {\cal
O} (N_c^{-1}) \label{ccs} \end{equation}
 This is precisely of the
form of eq.~(\ref{phen}) as required by large $N_c$ continuum
reduction.

A few  words about renormalization are in order.  In practice the
only viable way to numerically estimate the functional integral
for the partition function is via the lattice and one might worry
that using a lattice regularized theory might invalidate
eq.~(\ref{ccz}).  However, if one uses fermions which respect
chiral symmetry ({\it eg.} overlap fermions \cite{ol}) then one
can use eq.~(\ref{ccz}) to define a regulated chiral condensate in
a consistent manner.  As noted above the combination $m
\overline{q} q$ appears in the Lagrangian and as such has no
operator mixing and no anomalous dimension.  However, this does
not mean the expectation value is defined unambiguously.  Indeed,
even for a free field theory this quantity diverges in the
ultraviolet.  To render the operator finite it is typically
defined at some scale $\mu$ where fluctuations at scales above
$\mu$ are subtracted off.  Provided $\mu$ is large enough this is
computable perturbatively. Moreover in the regime
$\Lambda>>\mu>>T$, where $\Lambda$ is the cutoff scale of the
regularized theory, ({\it eg.} $1/a$ for a lattice regularized
theory) then eq.~(\ref{ccz}) can be used to compute $m \langle
\overline{q}q\rangle_T - m \langle \overline{q}q \rangle_{T=0}$
and the result is independent of the choice of $\mu$.  The reason
for this is simple---the same high momentum fluctuations are
subtracted from both the finite $T$ and $T=0$ cases so that their
difference does not depend on the subtraction point. Thus, the
numerator of eq.~(\ref{ccs}) is independent of $\mu$ and of order
$N_c^0$.  The denominator {\it does} depend on $\mu$ but for any
value of $\mu$, it is of order $N_c^1$.  This yields the scaling
in eq.~(\ref{ccs}) for any $\mu$. Finally it should be noted that
in the chiral limit, the ratio in eq.~(\ref{ccs}) is independent
of $\mu$; both the numerator and denominator have
$\mu$-independent contributions at leading chiral order $m$ with
all $\mu$ dependence coming in at order $m^2$ or higher.

It is very easy to understand the origin of eq.~(\ref{ccs}).
Equation (\ref{simple}) has a very simple interpretation.  It
represents the amount each hadron contributes the spatially
integrated operator $m \overline{q} q$ weighted by the density of
the hadrons.  Clearly the contribution from a single meson is
order $N_c^0$ as the operator is a one-body quark operator and the
number of quarks in a meson is independent of $N_c$.  Since the
density of mesons is also of order unity one sees that the shift
in the condensate from $T=0$ is of order unity. In contrast the
condensate scales with the number of colors since it effectively
counts the number of active quarks species.  The scaling of
eq.~(\ref{ccs}) follows from this.  From this qualitative
discussion, it should be clear that  although the preceding
derivation was for the chiral condensate, the same basic feature
should occur for generic local Lorentz scalar operators.  The
expectation value at $T=0$ will be altered at finite $T$ due to
the contributions from non-interacting mesons and glueballs and
such contributions are characteristically smaller in $N_c$
counting.  Of course, the issues of operator renormalization for
more complicated operators becomes more involved.

The temperature independence is only supposed to hold below the
phase transition; if the particles don't interact one might worry
that no phase transition can occur.  However, as $N_c \rightarrow
\infty$ QCD has an infinite number of narrow mesons and glueball
and the number of accessible states below some energy $E$ can grow
exponentially with $E$.  If the number of accessible meson and
glueball states below $E$ goes as $e^{-R/T_c}$, then the total
density of mesons and glueballs $\sum_s \rho_s(T)$ will diverge
as $T \rightarrow T_c$. This is just the Hagedorn phenomenon and
its description predates QCD\cite{Hag}; presumably, the QCD phase
is increasing well described as a Hagedorn transition as $N_c
\rightarrow \infty$.  Of course, as the Hagedorn transition is
approached, the $1/N_c$ suppression in meson-meson interactions is
overwhelmed by an enhancement due to the diverging density; one
can no longer neglect interactions and the derivation presented
here breaks down.  Thus at or above $T_c$, nothing prevents the
chiral condensate (or other observable) from differing from its
$T=0$ value  at large $N_c$.

The phenomenological demonstration that below $T_c$ the chiral
condensate (and other observables) are independent of $T$ at
large $N_c$ can be used to derive the full result of continuum
reduction. Note, that virtually nothing in the derivation
depended on the spatial size being infinite. If one repeated the
derivation for a the system  quantized in a box with sides of
length $L_x$,$L_y$ and $L_z$ and standard boundary conditions
(periodic for bosons, anti-periodic for fermions), the only
alterations would be in eqs.~(\ref{den}) and (\ref{G}) for which
the following substitutions would be made:
\begin{eqnarray}
&{}&\int \frac{{\rm d}^3 k }{(2 \pi)^3} \rightarrow
\sum_{n_x,n_y,n_z}\frac{1}{L_x L_y L_z} \nonumber \\
 &{}&k^2
 \rightarrow  \left( \frac {2 \pi n_x}{L_x} \right)^2 + \left(
\frac {2 \pi n_y}{L_y} \right)^2 + \left( \frac {2 \pi n_z}{L_z}
\right)^2 \; .
\end{eqnarray}
The key point is that with this substitution eq.~(\ref{simple})
continues to hold as do the scaling rules in eq.~(\ref{sigscale})
and thus so does eq.~(\ref{ccs}).

Now imagine one starts with a box of infinite extent in all four
directions.  Labeling one of these directions as time one can use
the arguments leading to eq.~(\ref{ccs}) to reduce the length of
this dimension without altering the chiral condensate (to leading
order in $1/N_c$) providing we do not pass through the phase
transition; this is just a temperature shift.  Next one can
relabel the axes and denote another axis as the time direction
and reduce it. This also simply becomes a temperature shift on a
box which is now finite in one spatial dimension.  However, as
was just argued, temperature independence of the chiral
condensate holds for the case of boxes with finite sides. One can
similarly reduce the other two sides leading to the result that
at large $N_c$ the chiral condensate for an arbitrary size finite
box is equal to that of an infinite box provided no phase
transition is crossed. This is large $N_c$ continuum reduction.

The logic underlying this derivation of large $N_c$ continuum
reduction for the chiral condensate is complimentary to that of
ref.~\cite{cc}.  It is based only on well known phenomenology of
hadronic interactions and simple features of finite-temperature
field theory and thus avoids a reliance on assumptions about the
applicability of random matrix theory for this context. However,
the present argument is quite limited compared to that in
ref.~\cite{cc} in at least one critical way. In contrast to
ref.~\cite{cc} it provides no insight into why chiral symmetry is
broken in the first place; it merely explains why the chiral
condensate does not depend on box size (at large $N_c$) for boxes
larger than a critical size.

\acknowledgments The author acknowledges R.~Narayanan and H.~
Neuberger for helpful discussions.  This research was begun at
the ``Large N QCD'' workshop at the ECT* in Trento; the author
thanks the organizers of this workshop and also thanks the ECT*
for its the hospitality. This work was supported by the
U.S.~Department of Energy through grant DE-FG02-93ER-40762.


\begin{thebibliography}{99}
\bibitem{lncr1}J.~Kiskis, R.~Narayanan and  H.~Neuberger,
Phys.~Rev.~ {\bf D66} (2002) 025019.
\bibitem{lncr2} J.~Kiskis, R.~Narayanan and
H.~Neuberger,lat/0208060.
\bibitem{lncr3}R.~Narayanan and H.~Neuberger, Phys.~Rev.~Lett.~{\bf 91} (2003) 081601.
\bibitem{lncr4} J.~Kiskis,R.~Narayanan and H.~Neuberger, Phys.~Lett.~{\bf B574} (2003) 65.
\bibitem{EK}T.~Eguchi and H.~Kawai, Phys.~Rev.~Lett. {\bf} 48 (1982) 1063.
\bibitem{BHN} G.~Bhanot, U.~M.`Heller and H.~Neuberger, Phys.~Lett.~{\bf B113} (1982)
47.
\bibitem{cc} R.~Narayanan and H.~Neuberger, hep-lat/0405025.
\bibitem{chpt}  P.~Gerber and  H.~Leutwyler, Nucl.~Phys.~{\bf
B321}(1989) 387.
\bibitem{'tH} G.~`t Hooft, Nucl.~Phys.~{\bf B72} (1974)461.
\bibitem{Wit} E.~Witten, Nucl.~Phys.~{\bf B160} (1979)57.
\bibitem{ol}Herbert Neuberger, Phys.~Lett.~{\bf B417} (1998) 141;  Phys.~Lett.~{\bf B 427}
(1998)353.
\bibitem{Hag}R.~Hagedorn, Nuovo Cim.~Suppl.~{\bf 3} (1965) 147.
\end{thebibliography}
\end{document}